\documentclass[aps,preprint,showpacs,preprintnumbers,amsmath,amssymb]{revtex4}
\usepackage{graphicx}

\begin{document}

\preprint{USTC-ICTS-09-08}

\title{Microscopic description of the chiral Tomonaga-Luttinger liquid
at the fractional quantum Hall edge}

\author{Wei Huang}
 \email{weihuang@mail.ustc.edu.cn}
\author{Zhao-Long Wang}
 \email{zlwang4@mail.ustc.edu.cn}
\author{Mu-Lin Yan}
 \email{mlyan@ustc.edu.cn}
\affiliation{Interdisciplinary Center for Theoretical Study,
Department of Modern Physics\\
University of Science and Technology of China, Hefei, Anhui 230026, China}

\date{\today}

\begin{abstract}
The effective field theory of the fractional quantum Hall edge is reformulated
from microscopic dynamics. Noncommutative Chern-Simons theory is a microscopic
description for the quantum Hall fluid. We use it for reference. Considering
relabeling symmetry of the electrons and incompressibility of the fluid, we
obtain a constraint and derive a chiral Tomonaga-Luttinger liquid theory
containing interaction terms. We calculate one-loop corrections to the phonon
and electron propagators and get a new tunneling exponent. It agrees with
experiments.
\end{abstract}


\maketitle

\section{introduction}

The chiral Tomonaga-Luttinger liquid at the fractional quantum Hall edge
becomes an active subject since Wen's hydrodynamic and effective field
formulation \cite{Wen:1990se,Wen:1992vi,Wen:2004ym}.
The effective edge theory is derived from the bulk effective Chern-Simons theory.
It predicts a nonlinear current-voltage relationship $I \sim V^\alpha$ with
an universal exponent $\alpha$, e.g. $\alpha = 3$ at the filling fraction $\nu = 1/3$.
For the Jain fractions $\nu = n/(2n \pm 1)$, the power-law behavior is also predicted
by including the effect of residual disorder \cite{Kane:1994vb,Kane:1995}.

A number of experiments \cite{Chang:1996,Grayson:1998zz,Chang:2001}
establish the existence of Tomonaga-Luttinger-liquid-like behavior.
However, the tunneling exponent measured is different from the prediction,
e.g. $\alpha \approx 2.7$ at $\nu = 1/3$.
The discrepancy between experiment and theory has been addressed in
\cite{Chang:2003zze,Shytov:1998,Levitov:2001,Chamon:1994,Wan:2002,Wan:2003,
Yang:2003,Joglekar:2003,Orgad:2008,Mandal:2001,Zulicke:2003,Jolad:2009}.
Many works have attempted to explain the discrepancy.
For the compressible quantum Hall fluid, the edge density enhancement is considered
\cite{Shytov:1998,Levitov:2001}.
For $\nu = 1/3$ and other Jain fractions $n/(2n \pm 1)$, some papers
suggest that the discrepancy is due to edge reconstruction
\cite{Chamon:1994,Wan:2002,Wan:2003,Yang:2003,Joglekar:2003,Orgad:2008}.
In contrast, some propose that the exponent is not universal,
since the discrepancy persists even in the absence of edge reconstruction
\cite{Mandal:2001,Zulicke:2003,Jolad:2009}.
It's still an open question.

An elementary derivation of the Chern-Simons description of the quantum Hall
effect was given by Susskind \cite{Susskind:2001fb},
wherein he claimed that the noncommutative version of the description
is exactly equivalent to the Laughlin theory.
As many successes on the connection between two theories
have been achieved in a collection of papers
\cite{Susskind:2001fb,Polychronakos:2001mi,Hellerman:2001rj,
Karabali:2001,Cappelli:2006wa},
we would stress that noncommutative Chern-Simons theory
is a workable microscopic description for the quantum Hall fluid.
Recently, this theory has been extended for constructing
the hierarchy of fractional quantum Hall states \cite{Wang:2008pea}.

In this paper, we try to pursue two questions:
whether the edge states in fractional quantum Hall effects could be
described by means of microscopic dynamics rather than an effective theory;
whether a more reliable and sounder exponent $\alpha$ could be derived.

The strategy is the following.
Based on Susskind's microscopic derivation \cite{Susskind:2001fb},
we will reformulate Wen's edge theory \cite{Wen:1992vi,Wen:2004ym}.
Firstly we give a constraint by considering microscopic dynamics:
relabeling symmetry of the electrons and incompressibility of the fluid.
The constraint should be obviously more natural than that given
by choosing a gauge-fixing condition in Wen's theory.
Secondly we solve the constraint exactly. It's amazing to find that
the solution as well as the action has a total differential form.
Finally we reduce the $2 + 1$ dimensional Chern-Simons theory to
an $1 + 1$ dimensional noncommutative chiral Tomonaga-Luttinger
liquid theory, which contains interaction terms expanding to
all orders in the noncommutative parameter $\theta$.
The commutative limit of it is Wen's theory.

Furthermore, as our theory contains interaction terms, it will predict
a new exponent and may provide a solution to the discrepancy mentioned above.
So we calculate one-loop Feynman diagrams caused by the interactions.
We notice the existence of a shortest incompressible distance
and impose an ultraviolet cutoff to evaluate the integrals.
Then we get one-loop corrections to the phonon and electron propagators.
The electron propagator still exhibits a power-law correlation,
but with a newly corrected prediction of the exponent
which is in good agreement with the experimental results.
This is a support of our derivation.

Briefly, we derive a noncommutative field theory
of the quantum Hall edge from microscopic dynamics.
We would claim that it should be sounder than existing effective theories.
Previously \cite{Wang:2008pea}, we have argued that the
Chern-Simons description of the edge excitations would receive
a natural explanation from the microscopic construction.
Recently, a similar subject was discussed incompletely \cite{Cabra:2008}.

The outline of the paper is as follows.
In Sec.~\ref{sec:nccs} we review the microscopic derivation
of Chern-Simons theory and get the constraint.
In Sec.~\ref{sec:edge} we derive the noncommutative
version of the chiral Tomonaga-Luttinger liquid theory.
In Sec.~\ref{sec:corr} we calculate the full phonon and
electron propagators and give the corrected exponent.
We conclude in Sec.~\ref{sec:con}.

\section{\label{sec:nccs}microscopic derivation of Chern-Simons theory}

We'll begin with a review of the microscopic derivation of the Chern-Simons
description of the quantum Hall effect \cite{Susskind:2001fb}.

Consider a two-dimensional electron system, the discrete electrons should be labeled
with a discrete index $\alpha$. Under the relabeling (or permutation) of the electrons,
$\alpha \rightarrow \alpha'=\alpha'(\alpha)$,
the real space coordinates and the Lagrangian remain invariant ($i = 1, 2$)
\begin{equation}
x_i^{\alpha}(t) = x_i^{\alpha'}(t) ,\quad \delta L = 0 .
\end{equation}

We can introduce a continuous space $y$, e.g. with a lattice $y_i^{\alpha} = x_i^{\alpha}(0)$,
and define the fluid fields $x_i(y_j, t)$ on it with values
\begin{equation}\label{x}
x_i(y_j, t)|_{y_j = y_j^{\alpha}} = x_i(y_j^{\alpha}, t) = x_i^{\alpha}(t) .
\end{equation}

As $y$ is just a continuum description replacing $\alpha$, we can naturally choose the coordinates
so that the electrons are evenly distributed in $y$ with a constant density $\rho_0$.
The relabeling symmetry of $\alpha$ is replaced by the area preserving diffeomorphism (APD) of $y$.

Assuming that the system is adiabatic so that short range forces lead to an equilibrium and the
potential is $\rho$ dependent ($\rho = \rho_0 |\partial y / \partial x|$ is the real space density),
in a background magnetic field $B$ we can write the Lagrangian as
\begin{equation}\label{l}
L = \int d^2 y \rho_0 [ \frac{m}{2} \dot{x}^2 - V(\rho)
+ \frac{e B}{2} \epsilon_{ab} \dot{x}_a x_b ] .
\end{equation}

Consider an infinitesimal transformation $y_i' = y_i + f_i(y)$,
which is APD if and only if (iff) $|\partial y' / \partial y| = 1$,
i.e. $f_i = \epsilon_{ij} \partial \Lambda (y) / \partial y_j$
with $\Lambda$ being an arbitrary function.
The $x$ coordinates and the Lagrangian transform as
\begin{equation}\label{dx}
\delta x_a = \frac{\partial x_a}{\partial y_i} f_i(y)
= \epsilon_{ij} \frac{\partial x_a}{\partial y_i}
\frac{\partial \Lambda}{\partial y_j} ,
\end{equation}
\begin{equation}\label{dl}
\begin{gathered}
\delta L = \frac{\partial L}{\partial \dot{x}_a} \delta \dot{x}_a
+ \frac{\partial L}{\partial x_a} \delta x_a \\
= \frac{d}{d t} (\frac{\partial L}{\partial \dot{x}_a} \delta x_a)
+ (- \frac{d}{d t} \frac{\partial L}{\partial \dot{x}_a}
+ \frac{\partial L}{\partial x_a}) \delta x_a = 0 .
\end{gathered}
\end{equation}

Besides the equation of motion
\begin{equation}\label{eom}
\frac{d}{d t} \frac{\partial L}{\partial \dot{x}_a}
- \frac{\partial L}{\partial x_a} = 0 ,
\end{equation}
we arrive at a conserved quantity and the constraints
\begin{equation}\label{cons}
g^{-1}(y) =
\begin{cases}
\frac{\partial}{\partial y_j}
(\epsilon_{ij} \dot{x}_a \frac{\partial x_a}{\partial y_i})
& \text{if $B$ is absent,} \\
\frac{1}{2} \epsilon_{ij} \epsilon_{ab}
\frac{\partial x_b}{\partial y_j} \frac{\partial x_a}{\partial y_i}
= |\frac{\partial x}{\partial y}|
& \text{if $B$ is strong,}
\end{cases}
\end{equation}
where $g(y)$ is an arbitrary time independent function. When the magnetic field
is strong, the kinetic term is dropped; $\rho(x,t) = \rho_0 g(y) = \rho(x,0)$.

In the strong magnetic field, the lowest Landau level dominates and the system
behaves as the quantum Hall fluid. Due to the Pauli exclusion principle,
the electrons are incompressible with a minimal area $2 \pi l_B^2$
($l_B = 1/ \sqrt{e B}$ is the magnetic length) \cite{Wen:2004ym}.
In the absence of vortices (no quasiparticle excitation), we assume that at $t=0$
the electrons are in equilibrium and uniformly occupy the minimal area.
So $\rho(x,0)$ is constant, $g(y) = const / \rho_0$ can be set to unity.
The constraint becomes
\begin{equation}\label{con}
1 = |\frac{\partial x}{\partial y}| .
\end{equation}

Consider a fractional quantum Hall fluid with a filling factor
$\nu = 1/(2n + 1)$, where $n$ is a positive integer.
It is also incompressible due to the interaction \cite{Laughlin:1983fy},
so we have the same constraint. Specially, the minimal area becomes $2 \pi l_B^2 / \nu$.
As $\rho_0 = \rho(x,0) = (2 \pi l_B^2 / \nu)^{-1}$, the factor $\nu = 2\pi \rho_0 / e B$
is truly the ratio of electrons to magnetic flux quanta.

We must stress that the constraint is derived from microscopic dynamics
by considering relabeling symmetry of the electrons and incompressibility
of the fluid. It is more exact, general and natural than that given
by choosing the gauge-fixing condition \cite{Wen:1992vi,Wen:2004ym}.

Consider small deviations from the equilibrium solution $x_i = y_i$,
\begin{equation}\label{a}
x_i(y, t) = y_i + \epsilon_{ij} \frac{A_j(y, t)}{2\pi \rho_0}
\equiv y_i + \theta \epsilon_{ij} A_j ,
\end{equation}
where the gauge transformation of $A_i$ under APD is
\begin{equation}\label{da}
\delta A_i = 2\pi \rho_0 \frac{\partial \Lambda}{\partial y_i}
+ \epsilon_{ab} \frac{\partial A_i}{\partial y_a} \frac{\partial \Lambda}{\partial y_b} .
\end{equation}

Substituting it and dropping total time derivatives gives the Chern-Simons action
\begin{eqnarray}\label{cs}
S &=& \frac{e B}{2} \int d t d^2 y \rho_0 \epsilon_{ij} \dot{x}_i x_j \nonumber\\
&=& \frac{1}{4 \pi \nu} \int d t d^2 y \epsilon_{ij} \dot{A}_i A_j .
\end{eqnarray}

Notice that $y$ space has a basic area quantum $\theta = 1/ (\nu e B) = l_B^2 / \nu$.
It means that $y$ space is noncommutative.
In \cite{Susskind:2001fb}, Eqs.~(\ref{con}), (\ref{da}) and (\ref{cs})
are recognized as first order truncations of a noncommutative Chern-Simons theory,
which is defined by the Lagrangian
\begin{equation}\label{ncl}
L_{NC} = \frac{1}{4 \pi \nu} \epsilon_{\mu\nu\rho}
( A_{\mu} \ast \partial_{\nu} A_{\rho}
+ \frac{2i}{3} A_{\mu} \ast A_{\nu} \ast A_{\rho} ) ,
\end{equation}
where $*$ represents the usual Moyal star-product defined in terms of
the noncommutative parameter $\theta $ \cite{Connes:1994yd,Douglas:2001ba}.
In the following, however, we go another way.
By expanding to higher order in $\theta$, we can also involve the noncommutativity
of $y$ space and capture the discrete character of the electron system.

The exact solution of $x_i(y,t)$ and $A_i(y,t)$ can be calculated from the constraint. As
\begin{eqnarray*}
1 &=& \frac{1}{2} \epsilon_{ij} \epsilon_{ab}
\partial_i (y_a + \theta \epsilon_{am} A_m)
\partial_j (y_b + \theta \epsilon_{bn} A_n) \\
&=& 1+ \theta \epsilon_{im} \partial_i A_m
+ \frac{1}{2} \theta^2 \epsilon_{ij}\epsilon_{mn} \partial_i A_m \partial_j A_n ,\\
0 &=& \epsilon_{ij} \partial_i ( A_j
+ \frac{1}{2} \theta \epsilon_{mn} A_m \partial_j A_n ) ,
\end{eqnarray*}
the condition for $A$ is that
\begin{equation}\label{cona}
A_j + \frac{1}{2} \theta \epsilon_{mn} A_m \partial_j A_n
= \partial_j \phi(y,t) ,
\end{equation}
where $\phi(y,t)$ is an arbitrary scalar field.
Expand $A_j = \sum_{n=0}^\infty \theta^n a^{(n)}_j$,
$\phi = \sum_{n=0}^\infty \theta^n \varphi^{(n)}$,
\begin{eqnarray*}
a^{(0)}_j &=& \partial_j \varphi^{(0)} ,\\
a^{(n)}_j &=& \partial_j \varphi^{(n)}
+ \frac{1}{2} \theta \epsilon_{ab} \sum_{m=0}^{n-1} \partial_j a^{(m)}_a a^{(n-1-m)}_b .
\end{eqnarray*}
Noticing that
\begin{eqnarray*}
A_j &=& \partial_j \sum_{l=0}^\infty \theta^l \varphi^{(l)}
+ \frac{1}{2} \theta \epsilon_{ab} \\
&& \partial_j \partial_a (\sum_{m=0}^\infty \theta^m \varphi^{(m)})
\partial_b (\sum_{n=0}^\infty \theta^n \varphi^{(n)}) + O(\theta^2) ,
\end{eqnarray*}
we can redefine $A_i = \sum_{n=0}^\infty \theta^n f^{(n)}_i$ with
\begin{equation}\label{f}
f^{(0)}_i = \partial_i \phi ,\quad
f^{(n)}_i = \frac{1}{2} \epsilon_{ab} \sum_{m=0}^{n-1} \partial_i f^{(m)}_a f^{(n-1-m)}_b .
\end{equation}

Substituting the exact solution into Eq.~(\ref{cs}) gives a noncommutative action
\begin{equation}\label{s}
S = \frac{1}{4 \pi \nu} \int d t d^2 y \sum_{n=0}^\infty \theta^n s^{(n)} ,
\end{equation}
where
\begin{equation}\label{sn}
s^{(n)} = \sum_{m=0}^{n} \epsilon_{ab} \partial_t f^{(m)}_a f^{(n-m)}_b .
\end{equation}

\section{\label{sec:edge}at the edge}

Since a two-dimensional electron gas on a quantum Hall plateau
is incompressible \cite{Laughlin:1983fy}, the edge excitations
are the only gapless excitations \cite{Halperin:1981ug}.
The edge states are important. We should study
whether the noncommutative action $S$ could describe them.
To describe the edge states, we need a one-dimensional theory.
To derive a one-dimensional theory, the first step is
to find a total differential form of the action.
It is difficult but has been done as follows.

Construct $F^{(n)}_\mu$ with total differentials
($\mu = 0, 1, 2$ and $\partial_0 \equiv \partial_t$):
$F^{(0)}_\mu = \partial_\mu \phi$ and for $n \geq 1$
\begin{eqnarray}\label{tdf}
F^{(n)}_\mu &=&
\partial_a ( \frac{1}{2} \epsilon_{ab} F^{(n-1)}_\mu F^{(0)}_b ) \nonumber\\
&-& \frac{1}{3} \sum_{m=1}^{n-1}
\partial_a ( \frac{1}{2} \epsilon_{ab} F^{(m)}_\mu F^{(n-1-m)}_b ) \nonumber\\
&+& \frac{1}{3} \partial_\mu ( \frac{1}{2} \epsilon_{ab} F^{(n-1)}_a F^{(0)}_b ) .
\end{eqnarray}

We find that $f^{(0)}_i = F^{(0)}_i$, $f^{(1)}_i = F^{(1)}_i$, etc.
Using Mathematica, the equivalence has been checked up to $n = 7$.
Logically, we make a conjecture: for all natural numbers $n$, $f^{(n)}_i = F^{(n)}_i$.

Similarly, every order of the Lagrangian density is a total differential,
$s^{(n)} = 2 F^{(n+1)}_0$. When the total time derivative is dropped,
\begin{equation}\label{tds}
s^{(n)} = \partial_a ( \epsilon_{ab} F^{(n)}_0 F^{(0)}_b )
- \frac{1}{3} \sum_{m=1}^{n} \partial_a ( \epsilon_{ab} F^{(m)}_0 F^{(n-m)}_b ) .
\end{equation}
We've checked it for $n \leq 6$ using Mathematica and for $n \leq 2$ by hand, e.g.
\begin{eqnarray*}
s^{(0)} & = & \epsilon_{ij} \partial_j [ \phi \partial_t \partial_i \phi ] ,\\
s^{(1)} & = & \frac{1}{3} \epsilon_{ij} \epsilon_{ab} \partial_j
[ \partial_t \partial_b \phi \partial_i \phi \partial_a \phi] ,\\
s^{(2)} & = & \frac{1}{4} \epsilon_{ij} \epsilon_{ab} \epsilon_{mn} \partial_j
[\partial_t (\partial_i \phi \partial_b \phi) \partial_a \partial_m \phi \partial_n \phi] .
\end{eqnarray*}

Being integration of a total differential, $S$ is nonzero and nontrivial iff a boundary exists.
Hence, we'll continue with a boundary, where some degrees of freedom become dynamical.
Consider the finite system $\Sigma$ confined by a simple potential well: an electric field $\vec{E}$.
The electrons drift in the direction perpendicular to $\vec{E}$ and $B$ and form an edge.
In the context of special relativity, in the frame $x$ moving with $v_i \equiv \epsilon_{ij} E_j / B$,
the electric field vanishes so that the electrons can be treated the same as that in bulk.
The real space $x^R$ is
\begin{equation}\label{edgex}
x_i^R = x_i + v_i t = y_i + \theta \epsilon_{ij} A_j + v_i t .
\end{equation}
Substituting it into the edge action and dropping total time derivatives gives
\begin{eqnarray}\label{edges}
S_\Sigma &=& \int_\Sigma d t d^2 y \rho_0
( \frac{e B}{2} \epsilon_{ij} \partial_t x_i^R x_j^R - e E_i x_i^R ) \nonumber\\
&=& \int_\Sigma d t d^2 y \rho_0 [ \frac{e B}{2} \epsilon_{ij}
\partial_t (x_i + 2 \epsilon_{ia} \frac{E_a}{B} t) x_j - e E_i x_i ] \nonumber\\
&=& \int_\Sigma d t d^2 y \rho_0 \frac{e B}{2} \epsilon_{ij} \dot{x}_i x_j = S .
\end{eqnarray}
It confirms that the electric field vanishes in the frame $x$ and the co-moving
coordinates $y$. So we can use the same Chern-Simons theory as in bulk.

Notice the relationship of the co-moving coordinates $y$ and the laboratory frame $y^R$
\begin{equation}\label{frame}
\begin{gathered}
y_i^R = y_i + v_i t ,\quad t^R = t ,\\
\partial_t = \partial_t^R + v_i \partial_i^R ,\quad \partial_i = \partial_i^R .
\end{gathered}
\end{equation}
In terms of $y^R$, the edge action acquires the form
\begin{eqnarray}\label{labedges}
S_\Sigma &=& \frac{1}{4 \pi \nu} \int_\Sigma d t^R d^2 y^R \epsilon_{ij}
(\partial_t^R + v_a \partial_a^R) A_i A_j \nonumber\\
&=& \frac{1}{4 \pi \nu} \int_\Sigma d t^R d^2 y^R \sum_{n=0}^\infty \theta^n s^{(n)} .
\end{eqnarray}

In the laboratory frame, ignoring $R$ for ease of notation,
choosing $\vec{E} = E \hat{y}_2$ and restricting the fluid to $y_2 \leq 0$ for convenience,
we can reduce the edge action to an $1 + 1$ dimensional chiral boson theory
\begin{eqnarray}\label{labchirals}
S_\chi &=& \frac{1}{4 \pi \nu} \int d t d y_1
\phi (\partial_t + v \partial_1) \partial_1 \phi + O(\theta) \nonumber\\
&=& \frac{1}{4 \pi \nu} \int d t d y_1 \sum_{n=0}^\infty \theta^n \chi^{(n)} ,
\end{eqnarray}
where $v = E / B$ and
\begin{equation}\label{labchi}
\chi^{(n)} = - F^{(n)}_0 F^{(0)}_1 + \frac{1}{3} \sum_{m=1}^{n} F^{(m)}_0 F^{(n-m)}_1 ,
\end{equation}
with redefined $F^{(0)}_0 = (\partial_t + v \partial_1) \phi$ and for $n \geq 1$
\begin{eqnarray}\label{labchiralf}
F^{(n)}_0 &=& \partial_a ( \frac{1}{2} \epsilon_{ab} F^{(n-1)}_0 F^{(0)}_b ) \nonumber\\
&-& \frac{1}{3} \sum_{m=1}^{n-1}
\partial_a ( \frac{1}{2} \epsilon_{ab} F^{(m)}_0 F^{(n-1-m)}_b ) \nonumber\\
&+& \frac{1}{3} (\partial_t + v \partial_1)
( \frac{1}{2} \epsilon_{ab} F^{(n-1)}_a F^{(0)}_b ) .
\end{eqnarray}

If we ignore the discrete character of the fluid, $\theta \propto l_B^2 \to 0$,
we get the commutative limit of our microscopic description, which coincides with
the phenomenological effective theory on the edge effect \cite{Wen:1992vi,Wen:2004ym}.

In fact, we get a noncommutative chiral Tomonaga-Luttinger liquid theory
which is one-dimensional and contains interaction terms. The hallmark feature of
Tomonaga-Luttinger-liquid-like behavior will be shown in the next section.
The dimensional reduction confirms that the only gapless excitations of
a two-dimensional incompressible quantum Hall fluid are the edge excitations.
We stress that interaction terms make things different: vertices
and loop Feynman diagrams emerge and correct the phonon propagator.

\section{\label{sec:corr}corrections to the phonon and electron propagators}

We'll calculate the loop corrections to the phonon and electron propagators with $S_\chi$.

Following Wen's hydrodynamic formulation \cite{Wen:1990se,Wen:1992vi,Wen:2004ym},
we have the commutation relation \cite{Floreanini:1987as}
\begin{equation}\label{quantum}
[\frac{1}{2\pi} \partial_1 \phi (y_1), \phi (y'_1)] = -i \nu \delta(y_1 - y'_1) ,
\end{equation}
and the electron operator (fermionic while $1/ \nu$ is odd)
\begin{equation}\label{eo}
\Psi \propto e^{i \frac{1}{\nu} \phi} ,\quad
\Psi(y_1) \Psi(y'_1) = (-1)^{\frac{1}{\nu}} \Psi(y'_1) \Psi(y_1) .
\end{equation}
The electron propagator can be calculated via the phonon propagator
\begin{equation}\label{ep}
\left< T\{ \Psi^\dag (y_1,t) \Psi (0) \} \right>
= \exp[\frac{1}{\nu^2} \left< \phi (y_1,t) \phi (0) \right>] .
\end{equation}

With the commutation relation and the equation of motion
$(\partial_t + v \partial_1) \partial_1 \phi = 0$,
we can calculate the retarded Green's function
\begin{eqnarray*}
D_R (y_1,t)
&=& \theta(t) \left< [\phi (y_1), \phi (0)] \right> ,\\
(\partial_t + v \partial_1) \partial_1 D_R (y_1,t)
&=& \partial_t \theta(t) \partial_1 \left< [\phi (y_1), \phi (0)] \right> \\
&=& -i 2\pi \nu \delta (t) \delta (y_1) ,\\
D_R (y_1,t)
&=& \int \frac{d^2 p}{(2\pi)^2} e^{- i (\omega_p t - p y_1)} \tilde{D}_R (p) ,\\
V_2(p) \equiv \tilde{D}_R (p)
&=& \frac{-i 2\pi \nu}{(\omega_p - v p) p} .
\end{eqnarray*}

\begin{figure}[hbtp]
\includegraphics[width=0.4\textwidth]{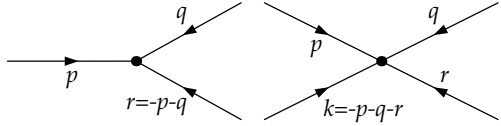}
\caption{\label{fig:v} 3-phonon and 4-phonon vertices.}
\end{figure}

To deal with $\partial_2$ in $\chi^{(n)}$ ($n \geq 1$),
we assume an undetermined distribution $\phi \propto \exp[h(y_2)]$.
Naturally, along the negative $y_2$ axis, $\exp[h(y_2)]$ should decrease
with a characteristic length $\sqrt{2 l_B^2 / \nu}$, which means the radius
occupied by every electron at the filling fraction $\nu$ \cite{Wen:2004ym}.
Using {\it Diagrammar} \cite{'tHooft:1973pz} with notations $y_\mu = (y_1 , i t)$,
$p_\mu = (p , i \omega_p)$ and $\int d^2 p = i \int d p d \omega_p$,
we can spell out the Feynman rules from the action times $i$ with the replacement
\begin{equation}\label{fourier}
\phi (y) = \int \frac{d^2 p}{(2\pi)^2}
\bar{\phi}(p) e^{i (p y_1 - \omega_p t)} e^{h(y_2)} .
\end{equation}
For $3$-phonon and $4$-phonon vertices (see Fig.~\ref{fig:v}),
the Feynman rules are ($w_p = \omega_p - v p$, the $\delta$ functions omitted)
\begin{eqnarray}\label{v3}
V_3(p,q) & \equiv &
\frac{\theta}{4\pi \nu} h' [ q(2p + q)w_p + p(p + 2q)w_q ] \nonumber\\
&=& \frac{-\theta}{4\pi \nu} h' \sum_{l=p,q,r} l^2 w_l ,
\end{eqnarray}
\begin{eqnarray}\label{v4}
V_4(p,q,r) & \equiv & \frac{i \theta^2}{4\pi \nu}
[ \frac{1}{4}(h'^2 + 2h'') \sum_{l=p,q,r,k} l^2 \sum_{l=p,q,r,k} l w_l \nonumber\\
&& - (h'^2 + h'') \sum_{l=p,q,r,k} l^3 w_l ] .
\end{eqnarray}

\begin{figure}[hbtp]
\includegraphics[width=0.4\textwidth]{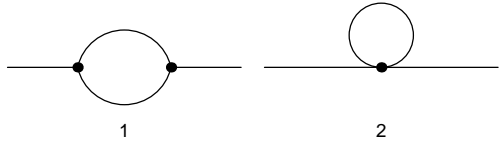}
\caption{\label{fig:g} One-loop Feynman diagrams.}
\end{figure}

Let $G(p)$ denote the sum of all 1PI (one particle irreducible) diagrams with two
external lines. We can express the full phonon propagator as
\begin{equation}\label{fullpp}
\begin{gathered}
\frac{-i 2\pi \nu}{(\omega_p - v p) p}
+ \frac{-i 2\pi \nu}{(\omega_p - v p) p} G \frac{-i 2\pi \nu}{(\omega_p - v p) p}
+ \cdots \\
= \frac{-i 2\pi \nu}{(\omega_p - v p) p + i 2\pi \nu G} .
\end{gathered}
\end{equation}
As shown in Fig.~\ref{fig:g}, the one-loop (second-order in $\theta$) contributions to $G(p)$ are
\begin{equation}\label{g1}
G_1 = \frac{1}{2} \int \frac{d^2 q}{(2\pi)^2} V_3(p,q) V_2(q) V_3(-p,-q) V_2(p+q) ,
\end{equation}
\begin{equation}\label{g2}
G_2 = \frac{1}{2} \int \frac{d^2 q}{(2\pi)^2} V_4(p,q,-p) V_2(q) ,
\end{equation}
where $G_n$ corresponds to the $n$th diagram in Fig.~\ref{fig:g}
($1/2$ is a symmetry factor).

Because $y$ space has a basic area quantum $\theta$ and a shortest incompressible
distance $l_B$, we can impose an ultraviolet cutoff $|q| \leq \Lambda$ and
$|\omega_q| \leq v \Lambda$ (due to the energy-momentum dispersion $\omega_q = v q$).
Via the uncertainty principle, $\Lambda = l_B^{-1}$.
To evaluate loop integrals, notice that: integrals over polynomials give zero,
$\int d q (q^2)^a = 0$, where $a$ is some nonnegative integer \cite{'tHooft:1973pz};
to the leading order
\begin{equation}\label{int}
\begin{gathered}
\int d q d \omega_q \frac{q}{\omega_q - v q} = -2 \Lambda^2 ,\\
\int d q d \omega_q \frac{q^2}{(\omega_q - v q + \omega_p - v p)(\omega_q - v q)}
= 2 \Lambda^2 \frac{1- \ln \Lambda}{v} .
\end{gathered}
\end{equation}
Then
\begin{eqnarray}\label{intg}
G_1 &=& \int \frac{d^2 q}{(2\pi)^2} \frac{\theta ^2}{8}
\frac{[q (2 p+q) w_p+p (p+2 q) w_q]^2 }{q w_q (p+q) (w_p+w_q)} h'^2 \nonumber\\
&=& - \frac{i}{(2\pi)^2} \theta ^2 \Lambda^2 h'^2
(p w_p + w_p^2 \frac{\ln \Lambda -1}{4 v}) ,\\
G_2 &=& \int \frac{d^2 q}{(2\pi)^2} \frac{\theta ^2}{4}
[ p^2 (h'^2+2 h'') -q^2 h'^2 \nonumber\\
&& + p w_p \frac{q^2 (h'^2+2 h'') - p^2 h'^2}{q w_q} ] \nonumber\\
&=& - \frac{i}{(2\pi)^2} \frac{1}{2} \theta ^2 \Lambda^2 (h'^2+2 h'') p w_p .
\end{eqnarray}
The full phonon propagator, to one-loop order, has the form
\begin{eqnarray*}
&& \frac{-i 2\pi \nu}
{p w_p [1+ \frac{\nu}{2\pi} \theta ^2 \Lambda^2 (\frac{3}{2}h'^2 + h'')]
+ w_p^2 \frac{\nu}{2\pi} \theta ^2 \Lambda^2 h'^2 \frac{\ln \Lambda -1}{4 v}} \\
&=& \frac{-i 2\pi \nu}{p w_p (1+ c_1) + w_p^2 c_2 \frac{\ln \Lambda -1}{v}} \\
&=& \frac{-i 2\pi \tilde{\nu}}{p (\omega_p - v p)}
- \frac{-i 2\pi \tilde{\nu}}
{p [ \omega_p - v p (1 - \frac{1+ c_1}{c_2} \frac{1}{\ln \Lambda -1}) ]} \\
&=& \frac{-i 2\pi \tilde{\nu}}{p (\omega_p - v p)}
- \frac{-i 2\pi \tilde{\nu}}{p (\omega_p - v_n p)} ,
\end{eqnarray*}
where
$c_1 = \frac{\nu}{2\pi} \theta ^2 \Lambda^2 (\frac{3}{2}h'^2 + h'')$,
$c_2 = \frac{\nu}{8\pi} \theta ^2 \Lambda^2 h'^2$,
$\tilde{\nu} = \frac{\nu}{1+ c_1}$ and
$v_n = v (1 - \frac{1+ c_1}{c_2} \frac{1}{\ln \Lambda -1})$.

Notice that: $\theta = l_B^2 / \nu$; $\Lambda = l_B^{-1}$;
$h'^2$ and $h''$ are proportional to $\nu / (2 l_B^2)$, because
$\exp[h(y_2)]$ decreases with the characteristic length $\sqrt{2 l_B^2 / \nu}$.
So $c_1$ and $c_2$ are constants independent of $\nu$ and $l_B$.
As a perturbation-theory correction should not be too large, we have $|c_1| \ll 1$.
Evidently $c_2 \geq 0$ so that $v_n$ is slightly smaller than $v$.
Because of the damping of the electric field caused by the presence of the
electrons, $v$ decreases by a small amount along the negative $y_2$ axis.
Without loss of generality we can choose $v_n$ to be the next-door neighbor
of $v$ and reconsider the full phonon propagator: the second term of the
propagator with $v$ cancels the first term with $v_n$, and so on;
the second term with $v_1$ can be ignored while it's non-chiral and cancels
the propagator in bulk with $v_0 = 0$; a sum over all slices gives
the overall full phonon propagator
\begin{equation}\label{overallpp}
\frac{-i 2\pi \tilde{\nu}}{p (\omega_p - v p)} .
\end{equation}

We can determine the undetermined distribution by solving
$c_1 = \frac{l_B^2}{2\pi \nu} (\frac{3}{2}h'^2 + h'')$.
As $\exp[h(y_2)]$ should decrease along the negative $y_2$ axis,
the only solution is
$h(y_2) = \sqrt{\frac{4\pi \nu}{3 l_B^2} c_1} y_2$ at $c_1 > 0$.
Naturally, we choose it and confirm the characteristic length
of this exponential distribution to be $\sqrt{2 l_B^2 / \nu}$.
Hence, $c_1 = 3/8\pi$ and $\tilde{\nu} = \frac{\nu}{1+ c_1} \approx 0.893 \nu$.
Then, as the position representation of the full phonon propagator is
\begin{equation}\label{ppp}
\left< \phi (y_1,t) \phi (0) \right> = - \tilde{\nu} \ln (y_1 - v t) + const ,
\end{equation}
the full electron propagator can be calculated as
\begin{equation}\label{fullep}
\left< T\{ \Psi^\dag (y_1,t) \Psi (0) \} \right> \propto \frac{1}{(y_1 - v t)^\alpha} ,
\end{equation}
where
\begin{equation}\label{alpha}
\alpha = \frac{\tilde{\nu}}{\nu^2} \approx 0.893 \frac{1}{\nu} .
\end{equation}
We see that the electron propagator at the fractional quantum Hall edge
exhibits a nontrivial power-law correlation, which indicates
Tomonaga-Luttinger-liquid-like behavior \cite{Chang:2003zze}.

At $\nu = 1/3$, the prediction of Eq.~(\ref{alpha}) is $\alpha \approx 2.68$.
It is in good agreement with the value measured in experiments
\cite{Chang:1996,Grayson:1998zz,Chang:2001}: $\alpha \approx 2.7$.

\section{\label{sec:con}conclusion}

In this paper, considering the microscopic dynamics of a two-dimensional
electron system in a strong perpendicular magnetic field, we have derived
a noncommutative field theory describing the chiral Tomonaga-Luttinger liquid
at the fractional quantum Hall edge. Without any adjustable parameter,
we have resolved the discrepancy of the exponent $\alpha$ between experiment
and the predictions of former effective field theories.

From the relabeling symmetry and the incompressibility of the fractional quantum
Hall system, we obtain a constraint. The constraint is more natural than that chosen
in \cite{Wen:1992vi,Wen:2004ym} and captures the discrete character of the system.
We solve the constraint and find a total differential form of the solution.
As also a total differential, the action is reduced to an $1 + 1$ dimensional chiral
Tomonaga-Luttinger liquid theory, which is the noncommutative version of Wen's theory
and contains interaction terms expanding to all orders in $\theta$.
Then one-loop corrections to the phonon and electron propagators are calculated.
The electron propagator exhibits a new power-law correlation, where the exponent
$\alpha$ is corrected to agree with experiments.

Furthermore, higher order corrections and the edge structures of hierarchial liquids
are remained as future subjects.

\begin{acknowledgments}
We wish to acknowledge the support of the grants from the NSF of China under
Grant No.~10588503, 10535060, 90403021, the grant from 973 Program of China
under Grant No.~2007CB815401 and the Pujiang Talent Project of the
Shanghai Science and Technology Committee under Grant No.~06PJ14114.
\end{acknowledgments}

\end{document}